\documentclass[fleqn,10pt]{wlscirep}
\usepackage[utf8]{inputenc}
\usepackage[T1]{fontenc}
\usepackage{graphicx}

\title{Scattering-lens based quantum imaging beyond shot noise}

\author[1,2]{Dong Li}
\author[1,2,*]{Yao Yao}
\affil[1]{Microsystems and Terahertz Research Center, China Academy of Engineering Physics, Chengdu Sichuan, 610200, China}
\affil[2]{Institute of Electronic Engineering, China Academy of Engineering Physics, Mianyang Sichuan, 621999, China}

\affil[*]{yaoyao\_mtrc@caep.cn}

\begin{abstract}
The scheme of optical imaging using scattering lens can provide a resolution beyond the classical optical diffraction limit with a coherent-state input. Nevertheless, due to the shot noise of the coherent state, the corresponding signal-to-noise ratio and resolution are both still shot-noise-limited. In order to circumvent this problem, we theoretically propose an alternative scheme where the squeezed state (with a sub-shot noise) is considered as input and the quantum noise is then suppressed below the shot-noise level. Consequently, when comparing with the previous imaging scheme (using combination of coherent state and scattering lens), our proposal is able to achieve an enhanced signal-to-noise ratio for a given scattering lens. Meanwhile, it is demonstrated that the resolution is also improved. We believe that this method may afford a new way of using squeezed states and enable a higher performance than that of using coherent state and scattering lens.
\end{abstract}
\begin{document}

\flushbottom
\maketitle
%
%
\thispagestyle{empty}

\section*{Introduction}

A significant fraction of the research activities in the field of high-resolution imaging involves the scattering lens \cite{van2011,choi2011,choi2014,park2014,yilmaz2015,hong2017,hong2018,leonetti2019,daniel2019,kanaev2018}. This is due to the fact that the optical imaging system using scattering lens could provide a better resolution than that of the conventional lens. Different from the conventional lens consisting of an ordered structure, the scattering lens is indeed a disordered medium \cite{wiersma2013,vellekoop2010np} which generally comprises randomly distributed small particles for light scattering. Previously, the disordered medium was deemed not suitable for the optical imaging. Since after a beam propagates through a disordered medium, it generally produces a speckle pattern owing to the multiple scattering, which was supposed to deteriorate the original information carried by the incident light. However, in recent years, it has been found that the disordered medium has the ability to overcome the classical diffraction limit \cite{van2011,choi2011} and can be used for the optical imaging \cite{choi2014,park2014}, even with a better performance than that of the conventional lens. In order to promote a more profound imagery, the disordered medium, utilized in the high-resolution imaging system, is usually called a scattering lens. 

In a traditional imaging configuration, it is mainly formed by a set of optical elements, in order from the object side: an input object, two conventional lenses, and a CCD camera. In contrast to the traditional scheme, many other methods with an extra scattering lens \cite{van2011,choi2011,choi2014,park2014} have been proposed for the high-resolution imaging in the past decades. According to the location of the scattering lens in the optical circuit, these schemes can be roughly divided into two categories: (I) behind the object \cite{choi2011,choi2014} (i.e. between the object and the conventional lens, corresponding to the wide-field imaging) and (II) in front of the object \cite{van2011} (in respect of the narrow-field imaging). It is worth pointing out that in case (I), the light behind the object would transport through the scattering lens while in case (II), the light illuminating the object is generated from the scattering lens. Although these two kinds of schemes show optical circuits with the different structures, both of them can realize an image with the enhanced resolution. This is because the scattering lens made of randomly distributed nanoparticles could increase the effective numerical aperture \cite{van2011,choi2011}.

\begin{figure}[tb]
\begin{center}
\includegraphics[width=.60\textwidth]{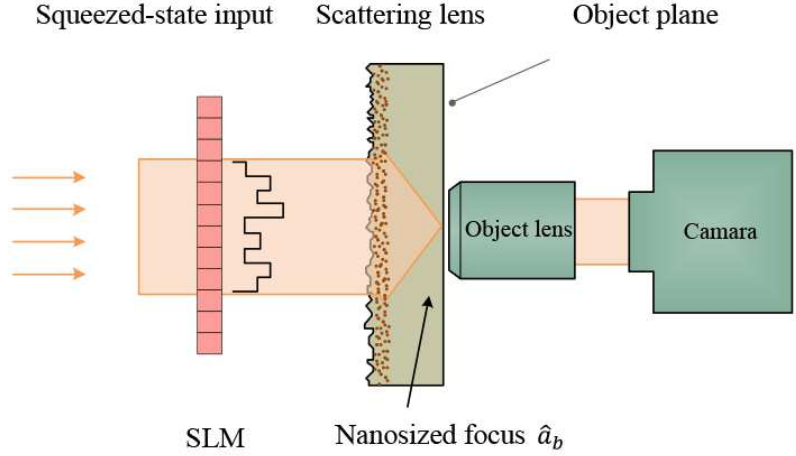}
\end{center}
\caption{Sketch of optical imaging scheme utilizing a scattering lens with the squeezed-state input. A scattering lens focuses a wavefront-shaped beam [modulated by the spatial light modulator (SLM)] on a small volume. The focus precisely illuminates a small object in the object plane. By scanning the focus and collecting the light behind the object, the image of the object can be obtained.}
\label{fig1}
\end{figure}

Particularly, we will concentrate on the case (II) in this work. In 2011, a high-resolution scheme of optical imaging using a scattering lens \cite{van2011} is proposed and experimentally realized as depicted in Fig. \ref{fig1}. In this scheme, the setup can be briefly described as follows: one wavefront-shaped coherent-state beam transports through a scattering lens and then produces a nanosized focus to illuminate a small object in the object plane. By scanning the nanosized focus and collecting the corresponding light behind the small object, the image of the object can be obtained. Importantly, this experiment yields some fantastic outcomes that the scattering lens with a {\it{coherent-state}} input can achieve a sub-$100$ nm resolution at visible wavelength, better than that of the conventional lens (an optimal resolution of order of around $200$ nm at visible wavelength \cite{van2011}).

Although this method provides a high resolution with a {\it{coherent-state}} input, the performance of imaging is still shot-noise-limited due to the shot noise of the coherent state (in fact, the performance of imaging is related to the quantum noise of the focused beam illuminating the object. When the coherent states are considered as input, the focused beam is still a coherent state \cite{li2019a}). To circumvent this problem, one possible way is to reduce the shot noise. Fortunately, the shot noise is actually a classical limit which can be beaten by the quantum technique \cite{caves1983,xiao1988,treps2002}.

As a typical nonclassical state, the squeezed state \cite{walls1983,lvovsky2015,andersen2016} is of significant importance since it possesses a quantum noise which can be below the shot noise level (note that a coherent state corresponds to the shot noise) \cite{walls2007,barnett2002,Beenakker2000}. As a consequence, the squeezed state can improve the signal-to-noise ratio (SNR) \cite{caves81,yurke86,xiao1987precision} and has been utilized in numerous applications, such as, quantum-enhanced magnetometer \cite{hor2012,ott2014}, gravitational wave detection \cite{aasi2013,barsotti2018,mehmet2018}, and quantum imaging using only conventional lenses (without any scattering lens) \cite{kolobov1993,kolobov1989,brida2010,sokolov2004,kolobov2000,b2005,chen2011}.

In order to achieve high performance, we propose an alternative scheme where the input is the \textit{squeezed state} instead of the coherent state, as shown in Fig. \ref{fig1}. In the presence of the squeezed-state input, the SNR of the optical imaging system is analyzed and the related resolution is also investigated. In addition, the comparison is performed between the squeezed and coherent states. It is found that the squeezed-state input leads to both an improved SNR and an enhanced resolution due to the suppressed quantum noise.

\section*{Results}
\subsection*{Propagation of the quantized light through a scattering lens}

\begin{figure*}[tbh]
\begin{center}
\includegraphics[width=.50\textwidth]{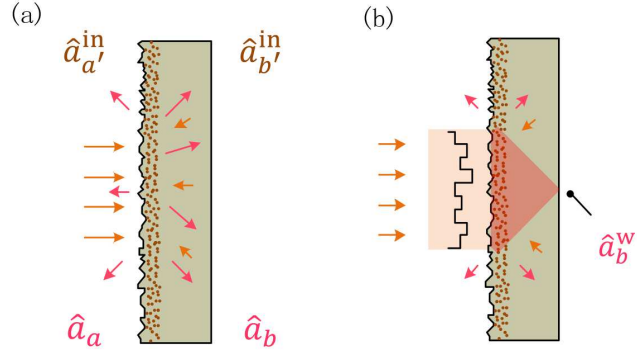} {}
\end{center}
\caption{Quantized light propagating through a scattering lens (a) in the absence of wavefront shaping, (b) in the presence of wavefront shaping. $\hat{a}_{a'}^{\rm{in}}$ ($\hat{a}_{b'}^{\rm{in}}$) represents the annihilation operator of the input mode and $\hat{a}_{a}$ ($\hat{a}_{b}$) the output mode. When the beams are injected, without the wavefront shaping in (a), the scattering lens separates the light into different optical channels randomly. As a result, the output presents a speckle pattern. In (b), with the wavefront shaping, the scattering lens couples the beams into the desired optical paths. Hence the output presents an ordered pattern. The wavefront shaping, performed by a spatial light modulator in (b), controls the phase of incident light.}
\label{fig2a}
\end{figure*}

Unlike the case of the conventional lens, the light focusing via a scattering lens \cite{tay2014,ojambati2016,fang2017,vellekoop2007,vellekoop2008,mosk2012} actually contains two indispensable processes: (I) multiple scattering of light inside the scattering lens and (II) shaping the wavefront of light before the light transporting through the scattering lens. As a matter of fact, wavefront shaping is an emerging technology for optical imaging and focusing through disordered media \cite{vellekoop2008,mosk2012,popoff2010}, by modulating the incident wavefront, which paves a way for manipulating the scattered light in an expected pattern. Generally, wavefront shaping can be performed by a spatial light modulator in experiment as shown in Fig. \ref{fig1}. The spatial light modulator acting as a reprogrammable matrix of pixels imprints desired phase values on the coherent wavefront.

Let us first review the process (I). Fig. \ref{fig2a}(a) depicts the propagation of quantized light through a scattering lens \cite{li2019,lodahl2006,an2018,xu2017a,xu2017b}, which comprises randomly distributed small particles for light scattering. To characterize the scattering lens, two primary factors are introduced: the transport mean free path $l$ and the thickness $L$. If $l \ll L$, the multiple scattering events would occur and result in a speckle pattern \cite{beenakker1997}. Hereafter we define $s \equiv L/l$ which determines the degree of disorder. 

After multiple scattering, the scattered mode $b$ \cite{lodahl2006} can be written as
\begin{equation}
\hat{a}_{b} = \sum_{a'}{t_{a'b}  \hat{a}_{a'}^{\rm{in}}} + \sum_{b'}{r_{b'b}  \hat{a}_{b'}^{\rm{in}}},
\label{creation}
\end{equation}
where $\hat{a}^{\rm{in}}_{a'}$ ($\hat{a}^{\rm{in}}_{b'}$) denote the annihilation operator of the incident modes $a'$ ($b'$) and obey the commutation relation $[\hat{O}, \hat{O}^{\dagger}]=1$ ($\hat{O} = \hat{a}^{\rm{in}}_{a'}, \hat{a}^{\rm{in}}_{b'}$). The transmission and reflection coefficients $t_{a' b}$ and $r_{b'b}$, subject to a constraint \cite{li2019a} $\sum_{a'} |t_{a'b}|^2 + \sum_{b'} |r_{b'b}|^2 = 1$, can be approximately regarded as complex Gaussian random variables \cite{goodman2015}. Accordingly, $t_{a'b} = \sqrt{T_{a'b}} e^{i\phi_{a'b}}$ and $r_{b'b} = \sqrt{R_{b'b}} e^{i\phi_{b'b}}$ where $\phi_{a'b}$ ($\phi_{b'b}$) is indeed uniformly distributed in the interval [$0,2\pi$] while $T_{a'b}$ and $R_{b'b}$ are the variables of Rayleigh distribution. In addition, the ensemble-averaged transmission and reflection coefficients are given by $\overline{T_{a'b}} = 1/(Ms)$ and $\overline{R_{b'b}} = (1-1/s)/M$ \cite{rossum1999,lodahl2006b}, where $M$ represents the number of transmission channels and the overline indicates the average over ensembles. It is easily seen that as the disorder strength $s$ increases, the average transmission coefficient $\overline{{T_{a'b}}}$ decreases. It is worth noting that Eq. (\ref{creation}) quantifies the very general input-output relation. The specific characteristics of the multiple scattering disordered medium are represented by the reflection and transmission coefficients. For instance, $t_{a'b}$ describes the coupling between the output mode $b$ and the input mode $a'$.

Second, consider process (II) wavefront shaping as depicted in Fig. \ref{fig2a}(b). In the presence of wavefront shaping, the scattered light can be directed towards a focus in any desired output mode with a scattering lens \cite{vellekoop2007}. Mathematically, the related input-output relation \cite{li2019a} can be characterized as
\begin{equation}
\hat{a}_{b}^{\rm{w}} = \sum_{a'}{|t_{a'b }|  \hat{a}_{a'}^{\rm{in}}} + \sum_{b'}{r_{b'b }  \hat{a}_{b'}^{\rm{in}}},
\label{creation2}
\end{equation}
where the superscript $w$ denotes wavefront shaping. In contrast to Eq. (\ref{creation}), the complex transmission coefficient $t_{a'b}$ is replaced by $|t_{a'b }| $ in Eq. (\ref{creation2}), which results from the fact that the phase modulator exactly compensates the phase retardation of each transmission channel in the scattering lens. 

In experiment, it can be realized for focusing light with a scattering lens via wavefront shaping under the current condition in laboratory nowadays, since the setting of phase modulation has been intensively investigated in theory and experiments over recent decades \cite{popoff2010tm,yoon2015,zhang2018,jang2018,moore2009}. However, comparing with previous works concentrating mainly on the enhanced intensity of the focused beam, we will focus on the quantum fluctuation of the focused beam.

According to Eq. (\ref{creation2}), the expectation value of photon number operator of the focused beam correspondingly arrives at
\begin{eqnarray}
\label{xp1}
\langle\hat{n}_b^{\rm{w}}\rangle &=&  \langle\hat{a}^{\rm{w} \dagger}_b \hat{a}^{\rm{w}}_b\rangle  \nonumber\\
&=& \sum_{a' a''}|t_{a'b}||t_{a''b}| \langle\hat{a}^{\rm{in} \dagger}_{a'} \hat{a}^{\rm{in}}_{a''}\rangle + \sum_{b' b''} r_{b'b}^{\ast} r_{b''b} \langle\hat{a}^{\rm{in} \dagger}_{b'} \hat{a}^{\rm{in}}_{b''}\rangle \nonumber \\
&&+ \sum_{a'b'}(|t_{a'b}|r_{b'b} \langle\hat{a}^{\rm{in}\dagger}_{a'} \hat{a}^{\rm{in}}_{b'}\rangle + h.c.),
\end{eqnarray}
where $\langle\hat{n}_k^{\rm{in}}\rangle = \langle\hat{a}_k^{\rm{in} \dagger} \hat{a}_k^{\rm{in}}\rangle$ ($k = a',b'$) and this expectation value is universal for any input state.


\subsection*{Signal-to-noise ratio}
\subsubsection*{Variance of the photon number of the focused beam}
Consider the squeezed states as input, $| \Psi^{\rm{in}} \rangle = [\hat{D}(\alpha)\hat{S}(\zeta)  |0\rangle]^{\otimes N}$, with $N$ being the number of input modes, $\hat{D}(\alpha) = e^{\alpha \hat{a}^{\dagger} - \alpha^{\ast} \hat{a}}$ the displacement operator, and $\hat{S}(\zeta) = e^{ (-\zeta\hat{a}^{\dagger 2} +\zeta^{\ast} \hat{a}^{2})/2 }$ the squeezing operator (the complex number $\alpha=|\alpha| e^{i \phi_{\alpha}}$ and the complex number $\zeta = g e^{i \phi_s}$ with the real number $g$ denoting the squeezing strength). For simplicity, we assume that the number of transmission channels is equal to the number of input modes $M=N$. Note that in our scheme, the input beam on the left-hand side of the scattering lens is the squeezed state whereas the one on the right-hand side is actually the vacuum state (i.e. $\langle \hat{n}^{\rm{in}}_{b'}\rangle = 0$).

The variance of operator $\hat{O}$ is defined as
\begin{eqnarray}
\label{var20}
\langle (\Delta \hat{O})^2 \rangle \equiv \langle \hat{O}^2 \rangle - \langle \hat{O} \rangle^2,
\end{eqnarray}
where $\hat{O} = \hat{n}_b^{\rm{w}}$. That is to say, to obtain the variance, it requires to compute $\langle \hat{n}_b^{\rm{w}} \rangle$ and $\langle (\hat{n}_b^{\rm{w}})^2 \rangle$.

The expectation value $\langle\hat{n}_b^{\rm{w}}\rangle$, according to Eq. (\ref{xp1}), can be obtained
\begin{eqnarray}
\label{var2a}
\langle \hat{n}_b^{\rm{w}} \rangle= \sum_{a'=1}^{M}T_{a'b} \sinh^2 g + \sum_{a'=1}^{M}\sum_{a''=1}^{M} |t_{a'b}||t_{a''b}||\alpha|^2,
\end{eqnarray}
where we present the derivation in Section Methods. Consider that $|\alpha|^2 \gg \sinh^2 g$, the second term in Eq. (\ref{var2a}) dominates (To the best of our knowledge, the maximum achievable value for the squeezing parameter is around $g \approx 1.5$ in experiment \cite{m2011} and correspondingly $\sinh^2 g \approx 4.53$. In contrast, the order of magnitude of $|\alpha|^2$ can be easily greater than that of $\sinh^2 g$ for a bright squeezed state in experiment \cite{p1994}. Therefore, it is reasonable to assume that $|\alpha|^2 \gg \sinh^2 g$.). As a result, Eq. (\ref{var2a}) is roughly equal to
\begin{eqnarray}
\langle \hat{n}_b^{\rm{w}} \rangle \simeq&  |\alpha|^2(\sum_{a'a''} |t_{a'b}| |t_{a''b}|).
\end{eqnarray}

According to the definition of variance, one can write the variance of operator $\hat{n}^{\rm{w}}_b$ as
\begin{eqnarray}
\label{var30}
\langle (\Delta \hat{n}^{\rm{w}}_b)^2 \rangle \equiv \langle (\hat{n}^{\rm{w}}_b)^2 \rangle - \langle \hat{n}^{\rm{w}}_b \rangle^2.
\end{eqnarray}
The corresponding variance of photon number is found to be
\begin{eqnarray}
\label{eq27}
\langle (\Delta \hat{n}_b^{\rm{w}})^2 \rangle &=& \sum_{a'=1}^{M} \sum_{a''=1}^{M} T_{a'b} T_{a''b} 2 \sinh^2g \cosh^2 g + \sum_{a'=1}^{M} \sum_{ b'=1}^{M} T_{a'b} R_{b'b} \sinh^2 g  \nonumber \\
&&+ |\alpha|^2 (\sum_{a'=1}^{M} \sum_{a''=1}^{M} |t_{a'b}| |t_{a''b}| ) [1 - \sum_{a'=1}^{M} T_{a'b}(1-e^{-2g})],
\end{eqnarray}
where we have set $\phi_{\alpha} = \phi_{s}=0$ and the detailed derivation is shown in Section Methods. Particularly, if $|\alpha|^2$ is sufficiently large for the third term of Eq. (\ref{eq27}) to dominate, the variance can be simplified to
\begin{eqnarray}
\label{eq002}
\langle (\Delta \hat{n}_b^{\rm{w}})^2 \rangle \simeq \langle \hat{n}_b^{\rm{w}} \rangle [1 - \sum_{a'=1}^{M} T_{a'b}(1-e^{-2g})].
\end{eqnarray}
From Eq. (\ref{eq002}), it is easy to find that the variance of photon number depends upon the value of the squeezing parameter $g$. Thus it pays to consider the two opposing limits $g=0$ and $g \approx 1.5 > 0$ (in experiment, $g \approx 1.5$ has been reported \cite{m2011}). First, consider $g = 0$ (i.e. the coherent-state input), one can rewrite Eq. (\ref{eq002}) as 
\begin{eqnarray}
\label{varcoh001}
\langle (\Delta \hat{n}_b^{\rm{w}})^2 \rangle = \langle \hat{n}_b^{\rm{w}} \rangle,
\end{eqnarray}
which is exactly as expected. Note that when the input is a coherent state, the focused mode is still a coherent state due to the linear optical process in the scattering lens (the variance of photon number of a coherent state is equal to its mean photon number). Next, we consider the opposing limit $g \approx 1.5$ (i.e. the squeezed-state input and $e^{-2g} \to 0$). In this situation, Eq. (\ref{eq002}) can be then approximately reduced to
\begin{eqnarray}
\label{eq003a}
\langle (\Delta \hat{n}_b^{\rm{w}})^2 \rangle \simeq \langle \hat{n}_b^{\rm{w}} \rangle [1 - \sum_{a'=1}^{M} T_{a'b}].
\end{eqnarray}
Comparing Eqs. (\ref{varcoh001}) and (\ref{eq003a}), one can find that the squeezed-state input leads to the reduction of quantum fluctuation of the focused beam. Moreover, from Eq. (\ref{eq003a}), it is easily seen that the degree of this reduction is related to a specific set of transmission coefficients of the scattering lens. 

For convenience, the Fano factor is introduced and defined as
\begin{eqnarray}
\label{ff}
F \equiv \langle (\Delta \hat{n}_b^{\rm{w}})^2 \rangle/\langle \hat{n}_b^{\rm{w}} \rangle.
\end{eqnarray}
When the input is a coherent state (i.e. $g=0$), by plugging Eq. (\ref{eq003a}) into Eq. (\ref{ff}), the Fano factor is found to be $F=1$. In contrast to the coherent-state input, we consider that the squeezed state is injected (with a large $g$, i.e. $e^{-2g} \to 0$). By inserting Eq. (\ref{eq003a}) into Eq. (\ref{ff}), it is easy to verify that the Fano factor of the focused beam is roughly equal to
\begin{eqnarray}
\label{fano01}
F \simeq 1 - \sum_{a'=1}^{M} T_{a'b},
\end{eqnarray}
which is smaller than that of the coherent state. This result reveals that the squeezed-state input can achieve a focused beam with the lower quantum noise than that of the coherent state.

By averaging over all disorder ensembles, according to Eq. (\ref{fano01}), the average Fano factor is found to be
\begin{eqnarray}
\label{fano002a}
\overline{F} \simeq 1 - \frac{1}{s},
\end{eqnarray}
where the overline means the average and $\overline{\sum_{a'}T_{a'b}} = N \overline{T_{a'b}} = 1/s$ is used \cite{lodahl2006}. From Eq. (\ref{fano002a}), one can see that with the increase of $s$, the average Fano factor increases monotonically which indicates that the output quantum noise increases. It is worth pointing out that this conclusion is suitable for the regime of the strong squeezing strength ($e^{-2g} \to 0$).

\begin{figure}[tb]
\begin{center}
\includegraphics[width=.850\textwidth]{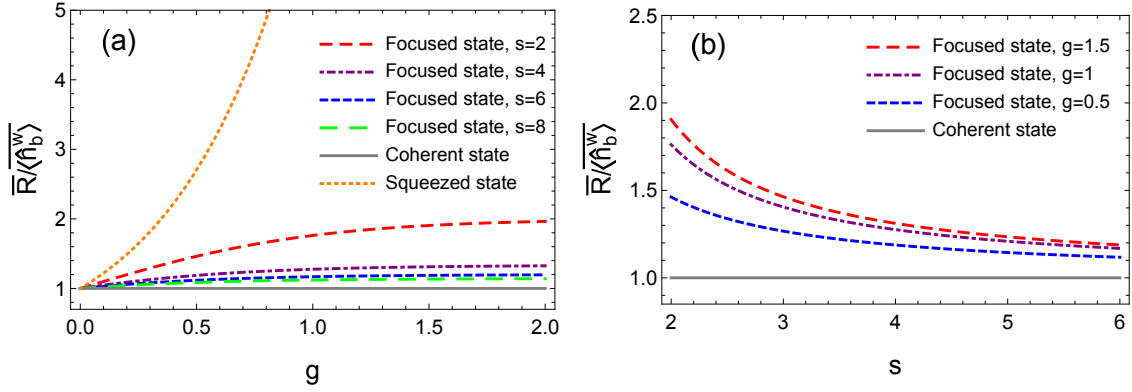} {}
\end{center}
\caption{The ratio of the average SNR $\overline{R}$ to the mean photon number $\overline{\langle \hat{n}^{\rm{w}}_{b} \rangle} $ as a function of (a) the squeezing parameter $g$ and (b) the disorder strength $s$ of the scattering lens. The input is the product state of $N$ squeezed states. Parameters: $|\alpha|^2=10000$.}
\label{figx}
\end{figure}

\subsubsection*{Signal-to-noise ratio}
In our scheme, The light illuminating the object for imaging is the focused beam as shown in Fig. \ref{fig1}. Accordingly, the focused beam determines the SNR of the imaging system \cite{b2005} which can be defined as
\begin{eqnarray}
\label{snr}
R \equiv \frac{\langle \hat{n}^{\rm{w}}_b \rangle^2}{\langle (\Delta \hat{n}^{\rm{w}}_b)^2 \rangle}= \frac{\langle \hat{n}^{\rm{w}}_{b} \rangle}{F} = \frac{\langle \hat{n}^{\rm{w}}_{b} \rangle}{1 - \sum_{a'=1}^{M} T_{a'b}(1-e^{-2g})},
\end{eqnarray}
where $\hat{n}^{\rm{w}}_b$ represents the photon number operator of the focused mode, $F$ denotes the corresponding Fano factor and $T_{a'b}$ characterizes the coupling between the output mode $b$ and the input mode $a'$. When $g=0$ (i.e. the coherent-state input and $F=1$), the SNR is naturally given by 
\begin{eqnarray}
\label{snrcoh001}
R = \langle \hat{n}^{\rm{w}}_{b} \rangle,
\end{eqnarray}
and the corresponding average SNR can be written as $\overline{R} = \overline{\langle \hat{n}^{\rm{w}}_{b} \rangle}.$ When $e^{-2g}\to 0$ (i.e. the squeezed-state input), the SNR can be written as
\begin{eqnarray}
\label{snr004a}
R = \frac{\langle \hat{n}^{\rm{w}}_{b} \rangle}{F} \simeq \frac{\langle \hat{n}^{\rm{w}}_{b} \rangle}{1 - \sum_{a'=1}^{M}T_{a'b}}.
\end{eqnarray}
By averaging over all the disorder ensembles, the ensemble-averaged SNR is obtained
\begin{eqnarray}
\label{snr004}
\overline{R} = \frac{\overline{\langle \hat{n}^{\rm{w}}_{b} \rangle}}{\overline{F}} \simeq \frac{\overline{\langle \hat{n}^{\rm{w}}_{b} \rangle}}{1 - 1/s},
\end{eqnarray}
where $\overline{\sum_{a'}T_{a'b}} = 1/s$ has been used and $s$ denotes the disorder parameter. From Eq. (\ref{snr004}), it is easy to check that the average SNR is improved as the decrease of disorder strength $s$ for a given $\overline{\langle \hat{n}^{\rm{w}}_{b} \rangle}$.

\begin{figure}[h]
\begin{center}
\includegraphics[width=.85\textwidth]{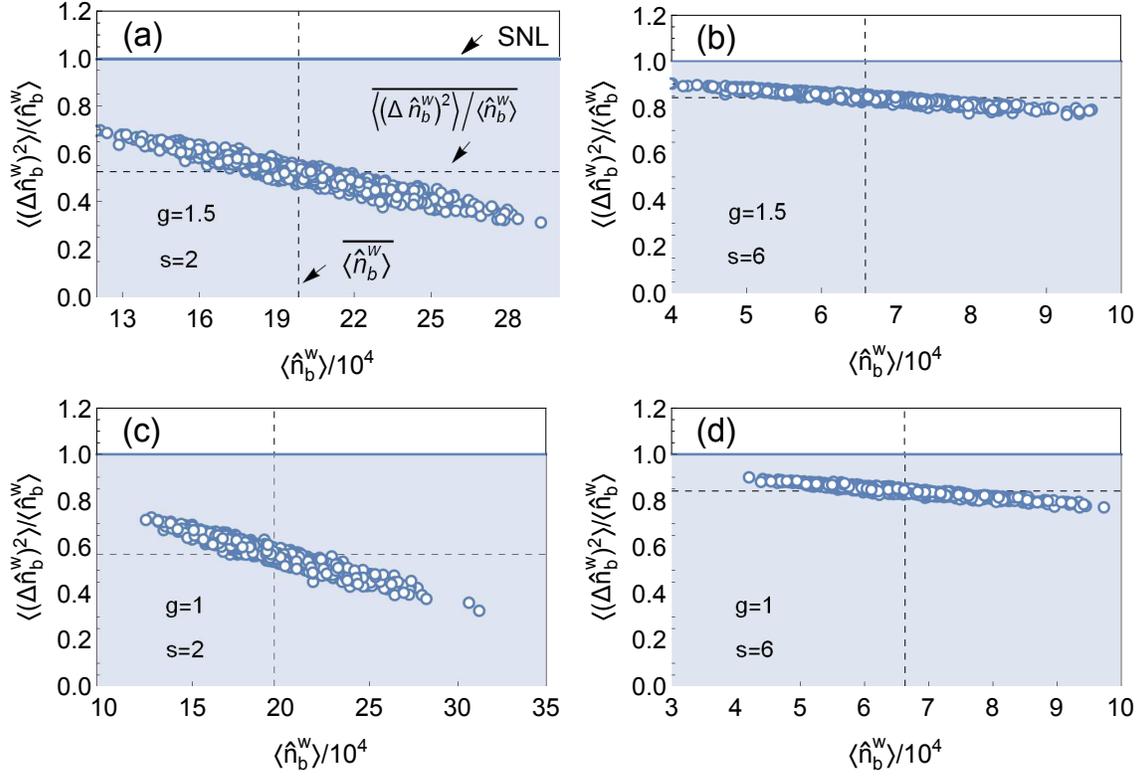} {}
\end{center}
\caption{Numerical simulation of Fano factor of the focused mode. Each point corresponds to one Fano factor. The numerical experiments have been repeated for 1000 times. In each simulation circle, the scattering matrix is generated randomly. The input beam is the photon-number squeezed state $| \Psi^{\rm{in}} \rangle = [\hat{D}(\alpha)\hat{S}(\zeta)  |0\rangle]^{\otimes N}$. Parameters: $|\alpha|^2 = 10000$, $N=50$.}
\label{figsamples}
\end{figure}

Based on Eq. (\ref{snr}), one can obtain the universal average SNR as $\overline{R} = \overline{\langle \hat{n}^{\rm{w}}_{b} \rangle} / [1 - (1-e^{-2g})/s]$. Figures \ref{figx}(a) and \ref{figx}(b) plot the average SNRs as a function of the squeezing parameter $g$ and the disorder strength $s$, respectively. It is obvious that the squeezed state always has a better average SNR than that of the coherent state. In other words, the squeezed state can improve the SNR. In Fig. \ref{figx}(a), with the increase of $g$, the average SNR increases which yields that the larger the squeezing parameter is, the better the average SNR is. However, in Fig. \ref{figx}(b), with the increase of $s$, this advantage resulting from the squeezed states would weaken due to the increased quantum noise induced from reflected modes (the vacuum state with the shot noise). It is worth noting that although the average SNR becomes worse, it is still better than that of the coherent state. To be more clear, we plot the numerical results of Fano factor for four cases (($g=1.5,s=2$), ($g=1.5,s=6$), ($g=1,s=2$), and ($g=1,s=6$)) in Fig. \ref{figsamples}. It is easily seen that in Fig. \ref{figsamples} the Fano factors are always smaller than one, which indicates the focused modes with a sub-shot noise under these four situations.

\subsection*{Resolution}

In addition to the SNR, the resolution is another significant criterion of performance evaluation for an optical imaging system. In this section, we would like to investigate how the squeezed states affect the resolution of imaging system. As a matter of fact, in 2005, Beskrovnyy and Kolobov \cite{b2005} developed a model that allows for analysis of the effects of squeezed state on the resolution. This model could be adopted in our circumstance. 

In essence, the analysis model is composed by two core parts: (i) the first step is to express the point spread function (PSF) \cite{b2005} in terms of prolate spheroidal functions \cite{xiao2001} within a quantum mechanical framework; (ii) the second one is to work out the ``cutoff'' PSF limited by the quantum noise (under the critical condition of the SNR decaying to one). Once the ``cutoff'' PSF is obtained, it is easy to find the corresponding resolution \cite{b2005} since the resolution is determined by the half-width of the ``cutoff'' PSF.

For the sake of clarity, we will not present the analysis model in this manuscript, since the detailed derivation is of extreme complexity involving a huge number of mathematical expressions (i.e. prolate spheroidal functions \cite{xiao2001}). Readers who are interested in more details about this model may refer to the Beskrovnyy and Kolobov paper \cite{b2005}. Based on this analysis model, the quantity that we consider is the super-resolution factor $J$ \cite{b2005} which will be briefly introduced in the following section.

\subsubsection*{Definition of the super-resolution factor}

In this section, we briefly introduce the definition of the super-resolution factor \cite{b2005}. The super-resolution factor, proposed by Beskrovnyy and Kolobov \cite{b2005}, is the ratio of the half width of PSF in quantum optics theory to the one in classical optics theory.

In classical optics theory, the PSF of the traditional imaging system \cite{b2005} is given by
\begin{eqnarray}
\label{eqh}
h(z-z') = \frac{\sin[c(z-z')]}{\pi(z-z')},
\end{eqnarray}
where $c$ denotes the spatial transmission bandwidth \cite{b1982} and $z$ ($z'$) is the spatial coordinate.

By contrast, within the quantum mechanical framework, the PSF of the modified imaging system \cite{b2005} is found to be
\begin{eqnarray}
h^{(r)}(z,z') = \sum_{k=0}^{Q-1} \phi_k(z) \phi_k(z'),
\end{eqnarray}
where the detailed derivation is shown in Section Methods, the real number $Q$ is determined by the SNR $R$ and $\phi_k(z)$ is the prolate spheroidal function \cite{b2005}. In fact, the set of functions $\phi_k(z)$ [$k = 1,2,3...$] constitutes a complete orthonormal basis \cite{xiao2001}.

To evaluate the resolution, the super-resolution factor $J$ \cite{b2005} is introduced and defined as
\begin{eqnarray}
J = \frac{W}{W_Q},
\end{eqnarray}
where we present the numerical analysis method to obtain $J$ in Section Methods, for brevity. $W$ [$W_Q$] is the half-width of the PSF $h(z-z')$ [$h^{(r)}(z,z')$].

\subsubsection*{Resolution}

\begin{figure*}[bt]
\begin{center}
\includegraphics[width=.40\textwidth]{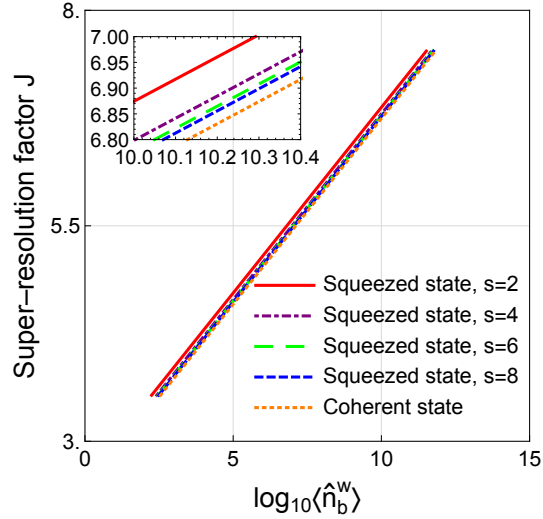} {}
\end{center}
\caption{Super-resolution factor $J$ as a function of the mean number of photon in the object plane. The red-solid, green-dashed, blue-dashed, orange-dotted curves denote the squeezed state ($s=2$), squeezed state ($s=4$), squeezed state ($s=6$), squeezed state ($s=8$), and coherent state, respectively. Parameters: $g=1.5$. }
\label{fig4}
\end{figure*}

Let us consider a typical condition for estimating the value of mean photon number in the focused beam illuminating the object in experiment. For instance, if it is given an incident beam with $\lambda = 694$ nm, optical power of $1$ mW and observation time of $1$ ms, the focused beam then has an optical power of $0.01$ mW (where it has been assumed that there is $1/100$ of total incident power in the focused mode \cite{vellekoop2007}). Finally, the mean photon number of the focused beam can be worked out as $\langle \hat{n}^{\rm{w}}_{b} \rangle \simeq 3.47 \times 10^{10}$.

Based on the order of value of $\langle \hat{n}^{\rm{w}}_{b} \rangle$ discussed above, we plot the super-resolution factors $J$ as a function of mean photon number $\langle \hat{n}^{\rm{w}}_{b} \rangle$ as shown in Fig. \ref{fig4}. In particular, the comparison between the squeezed-state input and the coherent-state one is performed. From Fig. \ref{fig4}, one can easily see that the squeezed state has a better resolution than that of the coherent state for a given intensity of the focused mode. In other words, the squeezed states could enhance the resolution owing to the sub-shot noise. We also investigate the effects of disorder strength $s$ on the resolution in Fig. \ref{fig4}. It can be seen that with the decrease of $s$, the resolution with squeezed state is improved for a fixed intensity $\overline{\langle \hat{n}^{\rm{w}}_b\rangle}$. This is due to the fact that the decrease of $s$ leads to the decrease of the average noise of the focused mode (Eq. (\ref{fano002a})). 

It is worth noting that the increase of the super-resolution factor is not dramatic as depicted in Fig. \ref{fig4}. Nevertheless, the squeezed state is still deserved to consider as input since any tiny improvement in high-precision measurement is very valuable.

\section*{Discussion}

\subsection*{Comparison between our proposal and the previous imaging scheme}
Our proposal is inspired by the scheme using coherent state with a scattering lens in the previous work \cite{van2011}. Therefore it is important to compare our proposal with the one in the reference \cite{van2011}. To be fair, we would like to consider the same situation (such as, same disorder parameter $s$). Before comparing these two schemes, let us estimate the value of $s$ in the former work \cite{van2011}. From the reference \cite{van2011}, the length of disordered medium is given by $L =2.88$ ${\rm{\mu m}}$. The mean free path is found to be $l =0.47 \pm 0.05$ ${\rm{\mu m}}$ according to the previous paper \cite{a1999} of A. Lagendijk (one of the authors of the reference \cite{van2011}). As a result, $s=L/l\approx 6$ in the former work \cite{van2011}. Accordingly, we also consider $s=6$ in our imaging scheme (see Figs. \ref{figx} and \ref{fig4}).

In Fig. \ref{figx}, it is easily found that the scheme using squeezed state (dashed-blue curve, $s=6$) has an enhanced SNR than that with coherent state (solid-gray curve), which results in an improved resolution as depicted in Fig. \ref{fig4}. That is to say, our scheme has a resolution beyond that in the reference \cite{van2011} when $s=6$ are same. Actually, this conclusion can be extended to the case of almost any value of $s$ (including not only weak scattering lens but also strong scattering lens).

It is worth noting that our claim on the improvement of imaging resolution is valid when comparing our proposal and the one in the reference \cite{van2011} under the same conditions (same parameter $s$). In fact, our analysis model does not involve the effect of $s$ on the increase of angular bandwidth of scattering lens $or$ the spatial size of the focused spot when analyzing the super-resolution factor $J$. This is still an open question. We would like to investigate this effect in the future.

\subsection*{SNR in the case of $N<M$}
Since the number of input squeezed-state modes $N$ may not be exactly equal to the number of transmission channels $M$ in experiment, it is worth discussing how this affects the SNR in imaging. In the situation of $N<M$, one can calculate the mean photon number
\begin{align}
\langle \hat{n}_{b}^{\rm{w}} \rangle_{N<M} &=\langle \hat{a}^{\rm{w} \dagger}_b \hat{a}^{\rm{w} }_b \rangle \nonumber \\
&=\sum_{a'=1}^{N} T_{a'b}\sinh^2 g+ \sum_{a'=1}^{N}\sum_{a''=1}^{N} |t_{a'b}| |t_{a''b}| |\alpha|^2.
\label{nbn}
\end{align}
Note that Eqs. (\ref{var2a}) and (\ref{nbn}) are different where $N<M$. Similar to Eq. (\ref{eq27}), the corresponding variance arrives at
\begin{align}
\langle (\Delta \hat{n}_{b}^{\rm{w}})^2 \rangle_{N<M}=& \sum_{a'=1}^{N} \sum_{a''=1}^{N} T_{a'b} T_{a''b} 2 \cosh^2 g \sinh^2 g + \sum_{a'=1}^{N} \sum_{b'=1}^{M} T_{a'b} R_{b'b} \sinh^2 g  + \sum_{a'=1}^{N} \sum_{a'' = N+1}^{M} T_{a'b} T_{a''b} \sinh^2 g \nonumber \\
& + |\alpha|^2 \sum_{a'=1}^{N} \sum_{a''=1}^{N} |t_{a'b}| |t_{a''b}| [ 1-\sum_{a'=1}^{N} T_{a'b}(1 - e^{-2g}),
\label{varnm}
\end{align}
where we have set $\phi_{\alpha} = \phi_{s} =0$. Compared to Eq. (\ref{eq27}), Eq. (\ref{varnm}) possesses an additional term of $\sum_{a'=1}^{N} \sum_{a'' = N+1}^{M} T_{a'b} T_{a''b} \sinh^2 g$. This term results from the interference between the squeezed state and the vacuum state from the empty input port.

\begin{figure*}[bt]
\begin{center}
\includegraphics[width=.850\textwidth]{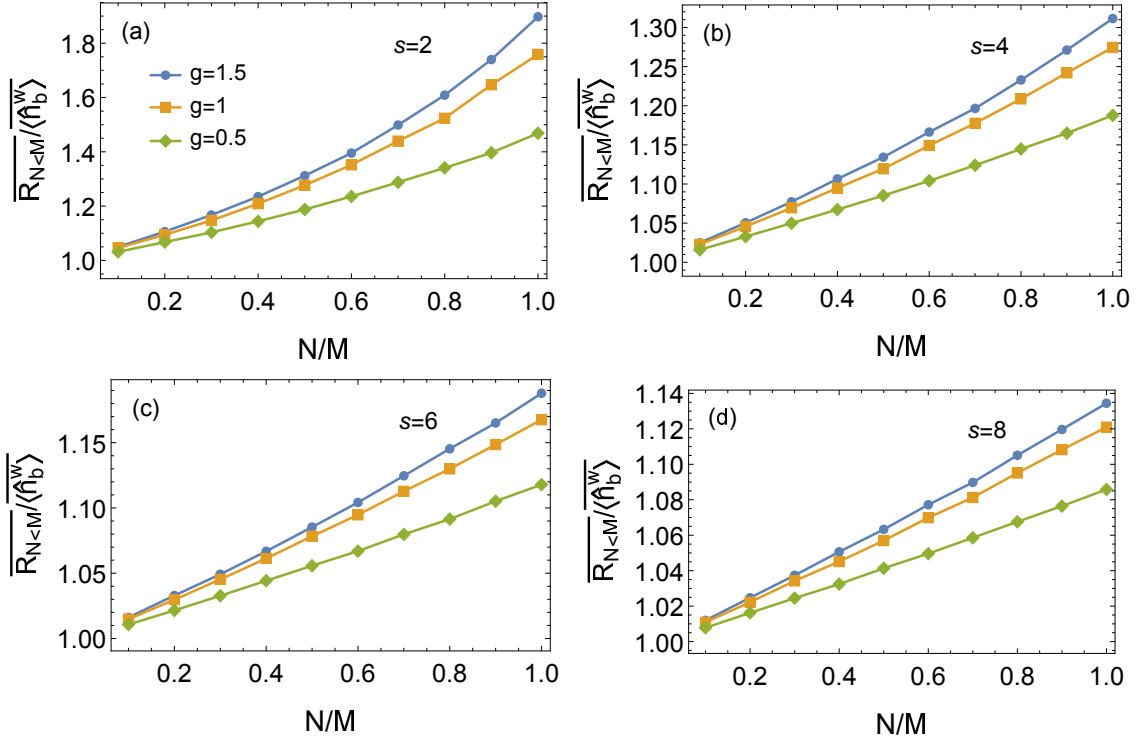} {}
\end{center}
\caption{The ratio of the average signal-to-noise ratio to the mean photon number $\overline{R_{N<M}}/\overline{\langle \hat{n}_{b}^{\rm{w}}\rangle}$ as a function of $N/M$. $N$: number of input modes, $M$: number of transmission channels. It is found that the increasing of $N/M$ leads to the increase in the ratio $\overline{R_{N<M}}/\overline{\langle \hat{n}_{b}^{\rm{w}} \rangle}$. Parameters: $|\alpha|^2=10000, M=50$.}
\label{fig00R}
\end{figure*}

By combining Eqs. (\ref{snr}), (\ref{nbn}), and (\ref{varnm}), one can obtain the signal-to-noise ratio $R_{N<M}$. The ratio $\overline{R_{N<M}}/\overline{\langle \hat{n}_{b}^{\rm{w}} \rangle}$ could be numerically analyzed. Fig. (\ref{fig00R}) plots the $\overline{R_{N<M}}/\overline{\langle \hat{n}_{b}^{\rm{w}} \rangle}$ as a function of $N/M$. From Fig. (\ref{fig00R}), it is easily found that the ratio $\overline{R_{N<M}}/\overline{\langle \hat{n}_{b}^{\rm{w}} \rangle}$ increases as the increasing of $N/M$. The maximum $\overline{R_{N<M}}/\overline{\langle \hat{n}_{b}^{\rm{w}} \rangle}$ can be achieved if $N/M=1$.

\subsection*{The output noise in the presence of $|\alpha|^2$ in various regimes }

\begin{figure}[htb]
\begin{center}
\includegraphics[width=.850\textwidth]{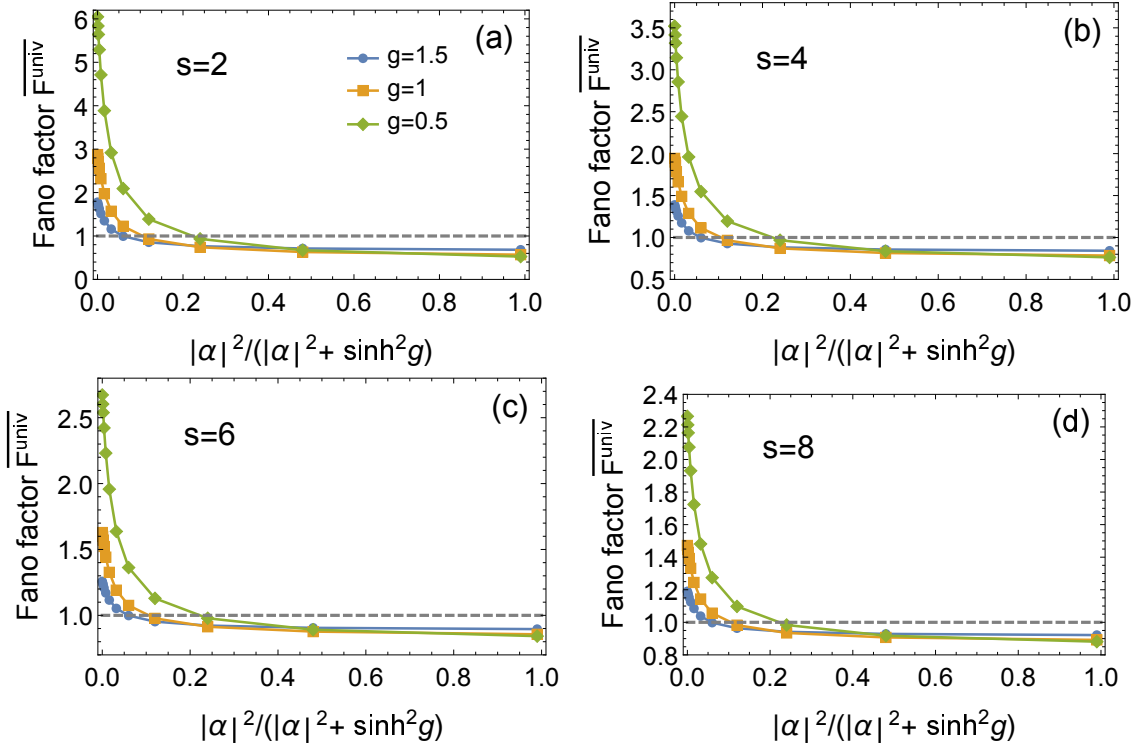} {}
\end{center}
\caption{The average Fano factor $\overline{F^{\rm{univ}}}$ of the focused mode as a function of $|\alpha|^2 / (|\alpha|^2 + \sinh^2 g)$, where $|\alpha|^2$ means the ``coherent'' intensity of input beam and $|\alpha|^2 + \sinh^2 g$ indicates the corresponding total intensity.}
\label{figfano}
\end{figure}

In contrast to the regime of large $|\alpha|^2$ in the previous section, we take into account a full range of $|\alpha|^2$. From Eqs. (\ref{var2a}), (\ref{eq27}) and (\ref{ff}), it is easy to compute the universal Fano factor $F^{\rm{univ}}$. Accordingly, the average $\overline{F^{\rm{univ}}}$ could be obtained by numerical simulation. Fig. \ref{figfano} presents the Fano factor $\overline{F^{\rm{univ}}}$ as a function of $|\alpha|^2 / (|\alpha|^2 + \sinh^2 g)$. It can be seen that with the increase of $|\alpha|^2 / (|\alpha|^2 + \sinh^2 g)$, the Fano factor $\overline{F^{\rm{univ}}}$ decreases. One can obtain the optimal $\overline{F^{\rm{univ}}}$ when $|\alpha|^2 / (|\alpha|^2 + \sinh^2 g) \simeq 1$. 

It is worth pointing out that when $|\alpha|^2 = 450$, the ratio $|\alpha|^2 / (|\alpha|^2 + \sinh^2 g) \simeq 0.99$ with $g=1.5$ ($\sinh^2 g \simeq 4.53$) which corresponds roughly to the optimal $\overline{F^{\rm{univ}}}$. Therefore, $|\alpha|^2 = 450$ could be considered as a strong coherent intensity when $g=1.5$, although it is not a strong intensity in experiment. Since $|\alpha|^2 > 450$ is easily achieved in experiment, it is reasonable to assume that $|\alpha|^2$ is in the regime of large value.

\subsection*{Effect of photon loss on the SNR }

\begin{figure}[tbh]
\begin{center}
\includegraphics[width=.8\textwidth]{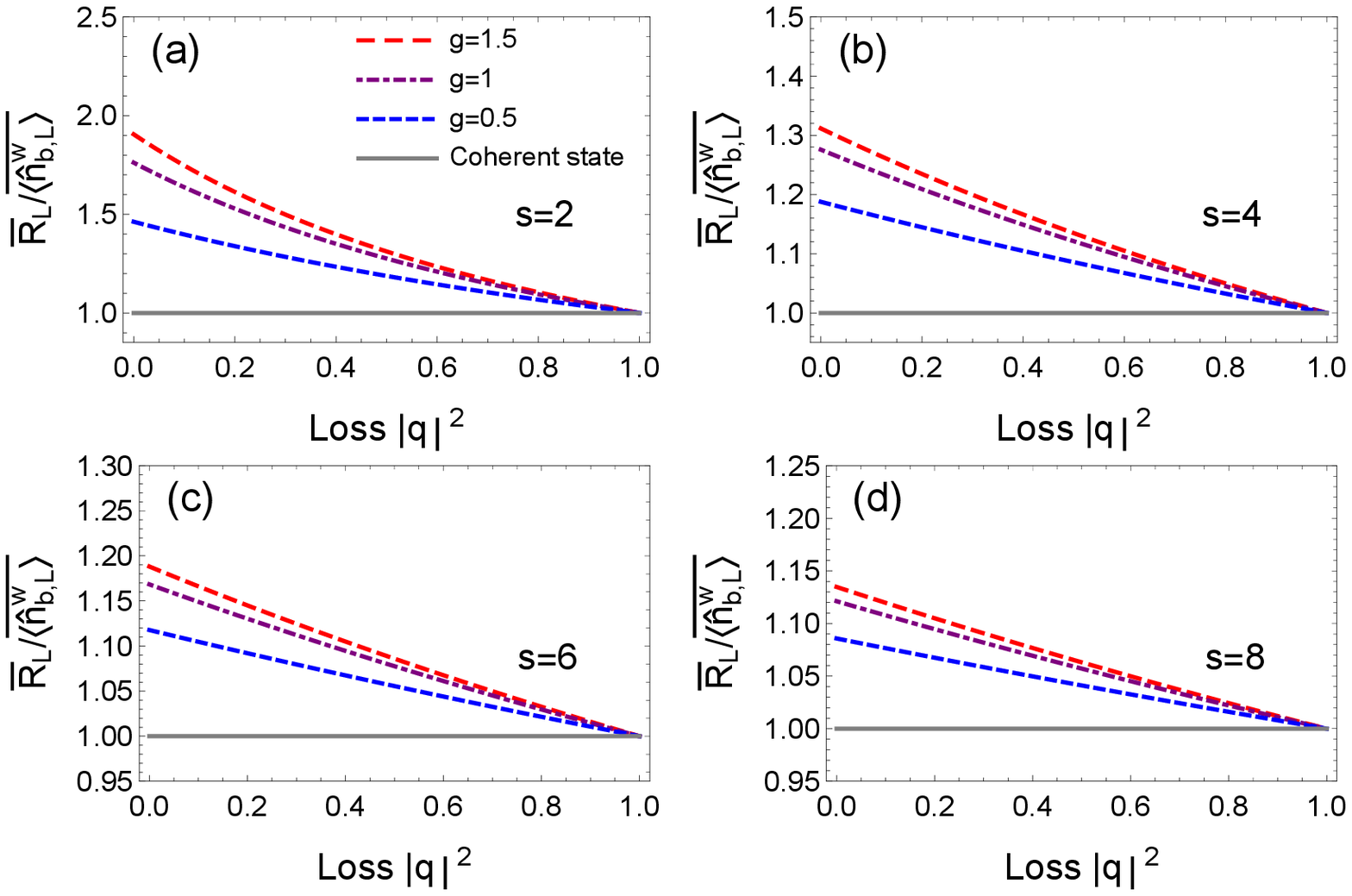} {}
\end{center}
\caption{The effect of photon loss on the average signal-to-noise ratio. Parameters: $|\alpha|^2 = 10000$, (a) $s=2$, (b) $s=4$, (c) $s=6$, (d) $s=8$.}
\label{figloss}
\end{figure}

In practical experiment, the photon loss is inevitable. Therefore, we consider the effect of photon loss of focused mode on the performance of imaging (SNR, particularly). To mimic the photon loss, we assume that the focused mode passes through a fictional two-port beam splitter with vacuum state injected in the other port. After propagating though the fictional beam splitter, the final output state can be characterized by
\begin{align}
\hat{a}^{\rm{w}\dagger}_{b,L} &= p^{\ast} \hat{a}^{\rm{w}\dagger}_{b} + q^{\ast} \hat{a}^{\rm{V}\dagger},\nonumber\\
\hat{a}^{\rm{w}}_{b,L} &= p \hat{a}^{\rm{w}}_{b} + q \hat{a}^{\rm{V}},
\label{eqnawbl}
\end{align}
where complex number $p$ and $q$ denotes transmission and reflection coefficients ($|q|^2$ indicates the photon-loss rate) and $\hat{a}^{\rm{V}}$ represents the annihilation operator of vacuum state.
The mean photon number of focus state with loss is found to be 
\begin{align}
\langle\hat{n}^{\rm{w}}_{b,L}\rangle= |p|^2 \langle\hat{n}^{\rm{w}}_{b}\rangle.
\end{align}
The average SNR is given by
\begin{align}
\overline{R_{L}}=\overline{\langle \hat{n}_{b,L}^{\rm{w}}\rangle}/\overline{F_{L}},
\end{align} 
where $\overline{F_{L}}$ can be expressed as
\begin{align}
\overline{F_{L}}= |p|^2 \overline{F} +|q|^2,
\end{align}
with $\overline{F}$ being the Fano factor without loss (For clarity, the detailed derivation is present in Section Methods). Fig. \ref{figloss} depicts the average SNR as a function of loss $|q|^2$ with various $g$. It is shown that although the averaged SNR degrades gradually as the increase of loss, the average SNR is still beyond the shot-noise level with small or moderate loss.

\section*{Conclusions}

Prior studies have shown that a high-resolution imaging can be achieved with a scattering lens via coherent states. This is owing to the fact that the scattering lens has a large numerical aperture and is able to focus the coherent-state light to a tighter spot than the diffraction limit of that conventional lens. Nevertheless, this is a pure classical technique without the help of any quantum technique. In the point of view of quantum optics, the squeezed state, as a nonclassical state, can enable a high-precision measurement. Therefore, in the current literature, we propose an alternative imaging scheme in which the squeezed states are considered as input instead of the coherent states.

Compared with the previous scheme, this new method takes advantage of not only the large numerical aperture of the scattering lens but also the sub-shot noise of the squeezed states. Consequently, our scheme with squeezed states has a better performance than that of the traditional imaging scheme. On the other hand, our scheme establishes a superiority that maintaining the same performance (i.e. resolution) requires a reduced photon number of the input states. This can eliminate the disadvantages due to the strong coherent-state input.

In summary, the effect of the squeezed-state input on the performance of imaging in the optical system with a scattering lens is investigated. It is clarified that the squeezed state leads to an improved signal-to-noise ratio in imaging compared with the coherent state due to the suppressed quantum noise. Moreover, it is found that the squeezed state also achieves an enhanced resolution in contrast to the coherent state. Therefore, our results may pave a new way to realize an image with a high resolution using scattering lens with the squeezed-state input.

\section*{Methods}

\subsection*{Variance of photon number of the focused mode}
\label{appd1}

Consider the squeezed states as input, $| \Psi^{\rm{in}} \rangle = [\hat{D}(\alpha)\hat{S}(\zeta)  |0\rangle]^{\otimes N}$, with $N$ being the number of input modes, $\hat{D}(\alpha) = e^{\alpha \hat{a}^{\dagger} - \alpha^{\ast} \hat{a}}$ the displacement operator, and $\hat{S}(\zeta) = e^{ (-\zeta\hat{a}^{\dagger 2} +\zeta^{\ast} \hat{a}^{2})/2 }$ the squeezing operator (the complex number $\alpha=|\alpha| e^{i \phi_{\alpha}}$ is the amplitude of the beam and the complex number $\zeta = g e^{i \phi_s}$ denotes the squeezing parameter, real number $g$ being the squeezing strength). The number of transmission channels is denoted by $M$. For simplicity, we assume that $M=N$.

For the convenience of calculation, one can rewrite the input-output relation of a disordered medium as
\begin{align}
\label{eq00a}
\hat{a}_b^{\rm{w}} = \sum_{a'} |t_{a'b}| (\hat{a}^{\rm{SV}}_{a'} + \alpha) + \sum_{b'} r_{b'b} \hat{a}_{b'}^{\rm{in}},
\end{align}
where the complex number $\alpha$ denotes the coherent amplitude of the input field and the operator $\hat{a}^{\rm{SV}}_{a'}$ accounts for the quantum fluctuation ($\langle \hat{a}^{\rm{SV}}_{a'} \rangle = 0$, $\langle 0| \hat{S}^{\dagger}(\alpha) \hat{D}^{\dagger}(\alpha) \hat{a}^{\rm{in}}_{a'}\hat{D}(\alpha)\hat{S}(\zeta) |0\rangle \to \langle 0| \hat{S}^{\dagger}(\alpha)  (\hat{a}^{\rm{SV}}_{a'} + \alpha )\hat{S}(\zeta) |0\rangle$).

From Eq. (\ref{eq00a}), it is easy to obtain the photon number operator
\begin{align}
\hat{n}_b^{\rm{w}} =& \sum_{a'=1}^{M} T_{a'b} \hat{a}^{\rm{SV} \dagger}_{a'}\hat{a}^{\rm{SV}}_{a'} + \sum_{b'=1}^{M} R_{a'b} \hat{a}^{\rm{in}\dagger}_{b'}\hat{a}_{b'}^{\rm{in}} + [\sum_{a'=1}^{M} \sum_{b'=1}^{M} |t_{a'b}| r_{b'b} \hat{a}^{\rm{SV} \dagger}_{a'}\hat{a}_{b'}^{\rm{in}} + H.c.] \\ \nonumber
& + \sum_{a'=1}^{M} \sum_{(a''\neq a',a''=1)}^{M} |t_{a'b}||t_{a''b}|\hat{a}^{\rm{SV} \dagger}_{a'}\hat{a}^{\rm{SV} }_{a''} + \sum_{b'=1}^{M} \sum_{(b''\neq b',b''=1)}^{M} r^{\ast}_{b'b} r_{b''b} \hat{a}^{\rm{in} \dagger}_{b'}\hat{a}_{b''}^{\rm{in}} \\ \nonumber
& + [\sum_{a'=1}^{M}|t_{a'b}| \alpha^{\ast} (\sum_{a''=1}^{M} |t_{a''b}| \hat{a}^{\rm{SV}}_{a''} + \sum_{b'=1}^{M} r_{b'b} \hat{a}_{b'}^{\rm{in}}) + H.c.] + |\alpha|^2\sum_{a'=1}^{M}\sum_{a''=1}^{M}|t_{a'b}||t_{a''b}| .
\end{align}
The expectation value of $\hat{n}_b^{\rm{w}}$ is then obtained
\begin{align}
\label{nb001}
\langle \hat{n}_b^{\rm{w}} \rangle =& \sum_{a'=1}^{M} |t_{a'b}|^2 \langle \hat{n}^{\rm{SV}}_{a'} \rangle + |\alpha|^2\sum_{a'=1}^{M} \sum_{a''=1}^{M} |t_{a'b}|  |t_{a''b}|,
\end{align}
where $\langle \hat{n}^{\rm{SV}}_{a'} \rangle = \langle \hat{a}^{\rm{SV} \dagger}_{a'}\hat{a}^{\rm{SV}}_{a'}\rangle = \sinh^2 g $ and we have used $\langle \hat{a}^{\rm{SV}}_{a'} \rangle = 0$ and $\langle \hat{a}_{b'} \rangle = 0$. Consider that $|\alpha|^2 \gg \sinh^2 g$, the second term in Eq. (\ref{nb001}) dominates. As a result, Eq. (\ref{nb001}) is roughly equal to
\begin{align}
\langle \hat{n}_b^{\rm{w}} \rangle \simeq&  |\alpha|^2\sum_{a'=1}^{M} \sum_{a''=1}^{M} |t_{a'b}|  |t_{a''b}|.
\end{align}

According to the definition of variance $\langle (\Delta \hat{n}^{\rm{w}}_b)^2 \rangle \equiv \langle (\hat{n}^{\rm{w}}_b)^2 \rangle - \langle \hat{n}^{\rm{w}}_b \rangle^2$, one can obtain the variance of photon number
\begin{align}
\label{eq27a001}
\langle (\Delta \hat{n}_b^{\rm{w}})^2 \rangle =& \sum_{a'=1}^{M} T_{a'b}^2 \langle (\Delta \hat{n}_{a'}^{\rm{SV}})^2 \rangle + \sum_{a'=1}^{M} \sum_{ b'=1}^{M} T_{a'b} R_{b'b} \langle\hat{a}^{\rm{SV} \dagger}_{a'} \hat{a}^{\rm{SV}}_{a'}\rangle \\ \nonumber
&+ \sum_{a'=1}^{M} \sum_{(a''\neq a',a''=1)}^{M} T_{a'b}T_{a''b}(\langle\hat{a}^{\rm{SV} \dagger}_{a'}\hat{a}^{\rm{SV}}_{a'}\hat{a}^{\rm{SV} \dagger}_{a''}\hat{a}^{\rm{SV}}_{a''} + \hat{a}^{\rm{SV} \dagger}_{a'}\hat{a}^{\rm{SV}}_{a'}\rangle + \langle\hat{a}^{\rm{SV} \dagger}_{a'}\hat{a}^{\rm{SV} \dagger}_{a'}\hat{a}^{\rm{SV} }_{a''}\hat{a}^{\rm{SV} }_{a''}\rangle) \\ \nonumber
& + |\alpha|^2 \sum_{a'=1}^{M}\sum_{a''=1}^{M} |t_{a'b}|  |t_{a''b}|  [\sum_{a'''=1}^{M}T_{a'''b} (2 \langle\hat{a}^{\rm{SV} \dagger}_{a'''} \hat{a}^{\rm{SV}}_{a'''}\rangle + 1)\\ \nonumber
& + \sum_{b'=1}^{M} R_{b'b} + (\sum_{a'''=1}^{M} |T_{a'''b}| \langle(\hat{a}^{\rm{SV}}_{a'''})^2\rangle + h.c.) ],
\end{align}
where $\langle (\Delta \hat{n}_{a'}^{\rm{SV}})^2 \rangle = \langle (\hat{n}^{\rm{SV}}_{a'})^2 \rangle -  \langle \hat{n}^{\rm{SV}}_{a'} \rangle^2$ and we have set $\phi_\alpha = 0$. Eq. (\ref{eq27a001}) could be further simplified to
\begin{align}
\label{eq27a002}
\langle (\Delta \hat{n}_b^{\rm{w}})^2 \rangle =& \sum_{a'=1}^{M}\sum_{a''=1}^{M} T_{a'b} T_{a''b} 2 \cosh^2g \sinh^2g + \sum_{a'=1}^{M} \sum_{ b'=1}^{M} T_{a'b} R_{b'b} \sinh^2g \\ \nonumber
& + |\alpha|^2 \sum_{a'=1}^{M}\sum_{a''=1}^{M} |t_{a'b}|  |t_{a''b}|  [1 - \sum_{a'''=1}^{M} T_{a'''b}(1-e^{-2g})],
\end{align}
where we have utilized $\sum_{b'} R_{b'b} = 1 -\sum_{a'} T_{a'b}$, $\langle  \hat{n}_{a'}^{\rm{SV}} \rangle = \sinh^2g$, $\langle (\Delta \hat{n}_{a'}^{\rm{SV}})^2 \rangle = 2 \cosh^2 g\sinh^2g$, $\langle\hat{a}^{\rm{SV} \dagger}_{a'}\hat{a}^{\rm{SV}}_{a'}\hat{a}^{\rm{SV} \dagger}_{a''}\hat{a}^{\rm{SV}}_{a''} + \hat{a}^{\rm{SV} \dagger}_{a'}\hat{a}^{\rm{SV}}_{a'}\rangle = \langle\hat{a}^{\rm{SV} \dagger}_{a'}\hat{a}^{\rm{SV} \dagger}_{a'}\hat{a}^{\rm{SV} }_{a''}\hat{a}^{\rm{SV} }_{a''}\rangle = \cosh^2g\sinh^2g$, $2 \langle\hat{a}^{\rm{SV} \dagger}_{a'} \hat{a}^{\rm{SV}}_{a'}\rangle + 1 + \langle(\hat{a}^{\rm{SV}}_{a'})^2\rangle + \langle(\hat{a}^{\rm{SV}\dagger }_{a'})^2\rangle = e^{-2g}$, and set $\phi_s =0$. Particularly, if $|\alpha|^2$ is sufficiently large for the fourth term of Eq. (\ref{eq27a002}) to dominate, the variance could be reduced to
\begin{align}
\label{eq002xx}
\langle (\Delta \hat{n}_b^{\rm{w}})^2 \rangle \simeq \langle \hat{n}_b^{\rm{w}} \rangle [1 - \sum_{a'=1}^{M} T_{a'b}(1-e^{-2g})],
\end{align}
where $\langle \hat{n}_b^{\rm{w}} \rangle \simeq |\alpha|^2\sum_{a'=1}^{M}\sum_{a''=1}^{M} |t_{a'b}|  |t_{a''b}|$ is used.

\subsection*{Quantum theory of optical imaging}

\subsubsection*{Input-output relation corresponding to a spatial Fourier transform}
Recall that in our imaging scheme, with the help of the squeezed states, the focused beam has a sub-shot noise. To uncover the role of the suppressed quantum fluctuation in the resolution of Fourier microscopy, it requires to analyze the configuration by using the quantum mechanical language.

\begin{figure*}[bth]
\begin{center}
\includegraphics[width=.60\textwidth]{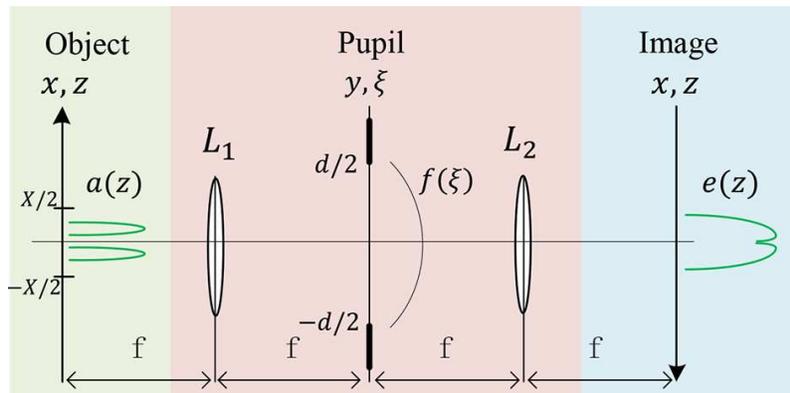} {}
\end{center}
\caption{The traditional imaging system, including the object field $a(z)$, the first lens ($L_1$), the field in the pupil plane $f(\xi)$, the second lens ($L_2$), and the field in the image plane $e(z)$. $X$ denotes the width of the object field and $d$ represents the finite size of pupil. For convenience, the dimensionless spatial coordinates $z = 2x/X$ in the object and image planes and $\xi = 2y/d$ in the pupil plane are introduced.}
\label{fig2lens}
\end{figure*}

In quantum optics, the object field in Fig. \ref{fig2lens} is described by the operator $\hat{a}(z)$ [$\hat{a}^{\dagger}(z)$] instead of the complex amplitude $a(z)$ in classical optics. Similarly, the field in the pupil plane is characterized by the operator $\hat{f}(\xi)$. The operators satisfy the standard commutation relation, $[\hat{a}(z), \hat{a}^{\dagger}(z')] = \delta(z - z')$ and $[\hat{f}(\xi), \hat{f}^{\dagger}(\xi')] = \delta(\xi - \xi')$. Since the lens $L_1$ performs a spatial Fourier transform between the object and pupil fields, the relation between $\hat{a}(z)$ and $\hat{f}(\xi)$ reads as follows \cite{b2005}
\begin{align}
\label{fainout001}
\hat{f}(\xi) = \sqrt{\frac{c}{2 \pi}} \int^{\infty}_{-\infty} \hat{a}(z) e^{-icz\xi} dz,
\end{align}
where $c$ denotes the spatial transmission bandwidth of the optical system.

\subsubsection*{Input-output relation in terms of the prolate spheroidal functions}
In terms of the prolate spheroidal functions, the operators $\hat{a}(z)$ and $\hat{f}(\xi)$ can be expressed as
\begin{align}
\label{azkzbz}
\hat{a}(z) = \sum_{k=0}^{\infty} \hat{a}_k \phi_{k}(z) + \sum_{k=0}^{\infty} \hat{b}_k \chi_k(z),\\
\hat{f}(\xi) = \sum_{k=0}^{\infty} \hat{f}_k \phi_{k}(\xi) + \sum_{k=0}^{\infty} \hat{g}_k \chi_k(\xi),
\label{fzfzgz}
\end{align}
where 
\begin{align}
	   \phi_k(z) =  \begin{cases}
	     \frac{1}{\sqrt{\lambda_k}} \psi_k(z) ,&  |z| \le 1, \\
	     0 ,&  |z| > 1,
	   \end{cases}
\quad \quad
	   \chi_k(z) =  \begin{cases}
	     0 ,&  |z| \le 1, \\
	     \frac{1}{\sqrt{1-\lambda_k}} \psi_k(z) ,&  |z| > 1,
	   \end{cases}
\end{align}
$\hat{a}_k$ ($\hat{b}_k$) and $\hat{f}_k$ ($\hat{g}_k$) denote the annihilation operators of the prolate mode $\phi_k$ ($\chi_k$) and can be obtained by 
\begin{align}
\label{abk001}
\hat{a}_k = \int_{-\infty}^{\infty} \hat{a}(z) \phi_k(z) dz,\\
\hat{b}_k = \int_{-\infty}^{\infty} \hat{a}(z) \chi_k(z) dz,\\
\label{abk002}
\hat{f}_k = \int_{-\infty}^{\infty} \hat{f}(\xi) \phi_k(\xi) dz,\\
\hat{g}_k = \int_{-\infty}^{\infty} \hat{f}(\xi) \chi_k(\xi) dz.
\end{align}
The operators ($\hat{a}_k$, $\hat{b}_k$, $\hat{f}_k$, and $\hat{g}_k$) obey the standard commutation relation, $[\hat{O}_k, \hat{O}_{k'}^{\dagger}] = \delta(k-k')$ ($O=a,b,g,k$). It is worthy noting that $\phi_k(z)$ lies within the region of $|z|<1$ while $\chi_k(z)$ is distributed over the area of $|z|>1$.

Based on the properties of the prolate spheroidal functions \cite{kolobov2007}, one has
\begin{align}
\label{properties001}
\int^{1}_{-1} \phi_k(z) e^{-i cz\xi} dz = (-i)^{k} \sqrt{\frac{2 \pi}{c}} \psi_k(\xi), \\
\int^{\infty}_{-\infty} \psi_k(z) e^{-i cz\xi} dz = (-i)^{k} \sqrt{\frac{2 \pi}{c}} \phi_k(\xi).
\label{properties002}
\end{align}

By substituting Eqs. (\ref{azkzbz}), (\ref{fzfzgz}), (\ref{properties001}), and (\ref{properties002}) into (\ref{fainout001}), it is easy to obtain the relations between the photon annihilation operators of the prolate modes in the object and pupil planes
\begin{align}
\label{eq38}
\hat{f}_k = (-i)^k (\sqrt{\lambda_k} \hat{a}_k + \sqrt{1-\lambda_k}) \hat{b}_k, \\
\hat{g}_k = (-i)^k (\sqrt{1-\lambda_k} \hat{a}_k - \sqrt{\lambda_k}) \hat{b}_k,
\label{eqs8a}
\end{align}
where Eqs. (\ref{eq38}) and (\ref{eqs8a}) build the connection between the input and output beams in the basis of the prolate spheroidal functions. Interestingly, this input-output relation is very similar to the case of a two-port beam splitter.

\subsubsection*{Reconstructed field operators and modified point-spread function}

From Eq. (\ref{eq38}), the operator-valued coefficients $\hat{a}_k^{(r)}$ of the reconstructed object \cite{b2005} is found to be
\begin{align}
\label{akr001}
\hat{a}^{(r)}_k = \frac{\hat{f}_k}{(-i)^k \sqrt{\lambda_k}} = \hat{a}_k + \sqrt{\frac{1-\lambda_k}{\lambda_k}} \hat{b}_k.
\end{align}

According to Eqs. (\ref{abk001}) and (\ref{akr001}), one can obtain the relation between the reconstructed field operator $\hat{a}^{(r)}(z)$ and the object field operator $\hat{a}(z)$ \cite{b2005}
\begin{align}
\label{ar002}
\hat{a}^{(r)}(z) = \int^{1}_{-1} h^{(r)}(z',z) \hat{a}(z') d z' + \sum^{Q-1}_{k=0} \sqrt{\frac{1-\lambda_k}{\lambda_k}}\hat{b}_{k}\phi_{k}(z),
\end{align}
where $h^{(r)}(z',z)$ denotes the reconstruction PSF and is given by
\begin{align}
\label{reconstr0}
h^{(r)}(z',z) = \sum_{k=0}^{Q-1} \phi_k(z') \phi_k(z).
\end{align}
From Eq. (\ref{reconstr0}), it is easily found that the modified imaging system has a reconstruction PSF which is related to the number $Q$. It is worthy pointing out that as $Q$ increases, the accuracy of PSF in Eq. (\ref{reconstr0}) is improved \cite{b2005}. Particularly, in the limit of $Q \to \infty$, the PSF is given by $h^{(r)}(z',z) = \lim\limits_{Q \to \infty }\sum_{k=0}^{Q-1} \phi_k(z') \phi_k(z) = \delta(z' - z)$  \cite{b2005}, which reveals that the image is a complete replication of the object. In other words, the PSF is totally accurate.

\subsection*{Super-resolution factor}

As discussed above, the model of input-output relation of the optical imaging system has been reviewed. According this relation, we will deal with the situation in our scheme particularly. Consider a point-like object placed at the origin $z=0$ in the object plane in Fig. \ref{fig1}. Correspondingly, we assume the focused beam in the object plane with a spatial distribution
\begin{align}
	   \langle \hat{a}^{\dagger}(z)\hat{a}(z) \rangle =  \begin{cases}
	     \frac{\langle\hat{n}^{\rm{w}}_b\rangle}{\varepsilon} ,&  |z| \le \varepsilon/2, \\
	     0 ,&  |z| > \varepsilon/2,
	   \end{cases}
\end{align}
where the width of the focused beam is very small $\varepsilon \sim 0$ and $\langle\hat{n}^{\rm{w}}_b\rangle$ is actually the mean photon number of the focused beam. It is easy to check that the total mean photon number of the beam illuminating the object is given by $\int_{-1}^{1} \langle \hat{a}^{\dagger}(z)\hat{a}(z) \rangle dz = \langle\hat{n}^{\rm{w}}_b \rangle$ and the corresponding reconstruction PSF is roughly equal to $h^{(r)}(0,z)$.

Assume that the object field (i.e. the focused beam) is a coherent state. In this situation, according to Eq. (\ref{snrcoh001}), the SNR is exactly equivalent to the mean photon number of the focused beam,
\begin{align}
R = \langle \hat{n}^{\rm{w}}_b \rangle = \int_{-1}^{1} \langle \hat{a}^{\dagger}(z)\hat{a}(z) \rangle dz.
\end{align}
On the contrary, by combining Eqs. (\ref{snr}), (\ref{ar002}), and (\ref{reconstr0}), the SNR of reconstructed object $R^{(r)}$ can be obtained,
\begin{align}
\label{reconstr}
R^{(r)} \equiv \frac{\langle \hat{n}^{(r)}  \rangle^2}{\langle (\Delta \hat{n}^{(r)})^2\rangle} = \left( \sum_{k=0}^{Q-1} |a_k|^2 \right)^2 \Biggm/ \left( \sum_{k=0}^{Q-1} |a_k|^2 / \lambda_k \right),
\end{align}
where the photon number operator of the reconstructed object $\hat{n}^{(r)} = \int_{-1}^{1} \hat{a}^{(r)\dagger}(z) \hat{a}^{(r)}(z) dz$ and $a_k = \int_{-1}^{1} \langle \hat{a}^{(r)}(z)  \rangle\phi_k(z) d z$ represents the coefficients of decomposition of $\hat{a}^{(r)}(z)$ over the prolate function $\phi_k(z)$.

Note that the number $Q$ in Eq. (\ref{reconstr0}) determines the PSF in the super-resolving-Fourier-microscopy imaging system. It is easily checked that increasing $Q$ improves the accuracy of the PSF in Eq. (\ref{reconstr0}) which is related to the resolution. The larger the number $Q$ is, the higher the super-resolution of the reconstructed object achieves. Nevertheless, from Eq. (\ref{reconstr}), it is easy to verify that with the increase of $Q$, the SNR in the reconstructed object degrades. That is to say, the number $Q$ could not be arbitrary large due to constraint from the decay of the SNR in the reconstructed object. Without loss of generality, one can presume that the SNR in the reconstructed object, no less than unity, can deliver the reconstruction of the object. 

\begin{figure*}[th]
\begin{center}
\includegraphics[width=.40\textwidth]{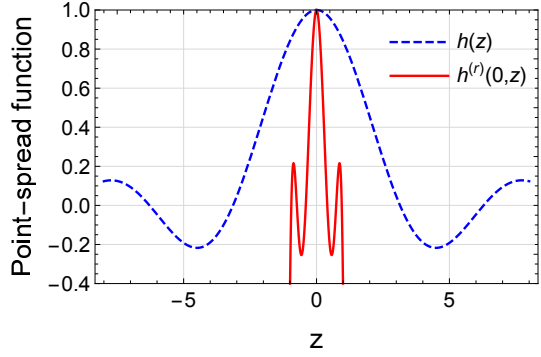} {}
\end{center}
\caption{Point-spread function as a function of $z$. The blue-dashed line denotes the traditional optical imaging system and the red one indicates the super-resolving-Fourier-microscopy imaging system.  }
\label{fig3}
\end{figure*}

To describe the superiority of this scheme in resolution, the comparison between the traditional and modified schemes is performed. Correspondingly, the super-resolution factor is introduced and is defined as the ratio of the width of the diffraction-limited imaging PSF ($W$) in Eq. (\ref{eqh}) to the one of the reconstruction PSF ($W_Q$). In order to characterize it more intuitively, Fig. \ref{fig3} plots the diffraction-limited imaging PSF $h(z)$ and the reconstruction PSF $h^{(r)}(0,z)$ for $Q=7$ normalized to unity at their maxima, respectively. To quantify the degree of the super-resolution, we introduce the half-widths $W$ and $W_Q$ of these two PSFs measured at their half maxima. Then the super-resolution factor $J$ can be obtained by
\begin{align}
J = \frac{W}{W_Q}.
\end{align}
As depicted in Fig. \ref{fig3}, it is easy to find that $W = 1.90$, $W_Q = 0.25$, and $J = 7.6$ which reproduces the result in Ref. \cite{b2005}. 

\subsection*{Derivation of the SNR in the presence of photon loss}

The detailed derivation of SNR with loss is shown as follows. Assume that the focus mode experiences photon loss. We mimic this loss process by passing through a fictional two-port beam splitter with vacuum state injected in the other input port. After the focus state passes through the beam splitter, the output mode could be described by
\begin{align}
\hat{a}^{\rm{w}}_{b,L} &= p \hat{a}^{\rm{w}}_{b} + q \hat{a}^{\rm{V}},
\label{eqnawbl}
\end{align}
where the complex number $p$ and $q$ denote transmission and reflection coefficients ($|q|^2$ indicates the photon loss rate), $\hat{a}^{\rm{V}}$ represents the annihilation operator of vacuum state.


Inserting Eq. (\ref{eq00a}) into (\ref{eqnawbl}), one can obtain
\begin{align}
\hat{a}^{\rm{w}\dagger}_{b,L} &= p^{\ast} \sum_{a'=1}^{M}|t_{a'b}^{\ast}| (\hat{a}^{\rm{SV} \dagger}_{a'} + \alpha^{\ast})  + p^{\ast}\sum_{b'=1}^{M} r_{b'b}^{\ast}\hat{a}_{b'}^{\dagger} + q^{\ast} \hat{a}^{\rm{V}\dagger},\nonumber\\
\hat{a}^{\rm{w}}_{b,L} &= p \sum_{a'=1}^{M}|t_{a'b}| (\hat{a}^{\rm{SV}}_{a'} + \alpha) + p\sum_{b'=1}^{M} r_{b'b}\hat{a}_{b'} + q \hat{a}^{\rm{V}},
\end{align}
The photon number operator can be cast into
\begin{align}
\hat{n}^{\rm{w}}_{b,L}=&\hat{a}^{\rm{w}\dagger}_{b,L}\hat{a}^{\rm{w}}_{b,L}\nonumber\\
 =&p^{\ast}q\hat{n}^{\rm{w}}_{b}+ q^{\ast} \hat{a}^{\rm{V}\dagger} [p \sum_{a'=1}^{M}|t_{a'b}| (\hat{a}^{\rm{SV}}_{a'} + \alpha) + p\sum_{b'=1}^{M} r_{b'b}\hat{a}_{b'}]+[p^{\ast} \sum_{a'=1}^{M}|t_{a'b}^{\ast}| (\hat{a}^{\rm{SV} \dagger}_{a'} + \alpha^{\ast})  + p^{\ast}\sum_{b'=1}^{M} r_{b'b}^{\ast}\hat{a}_{b'}^{\dagger}]q \hat{a}^{\rm{V}}.
\end{align}
The corresponding mean photon number is then given by
\begin{align}
\langle\hat{n}^{\rm{w}}_{b,L}\rangle=&\langle \hat{a}^{\rm{w}\dagger}_{b,L}\hat{a}^{\rm{w}}_{b,L} \rangle= |p|^2 \langle\hat{n}^{\rm{w}}_{b}\rangle,
\end{align}
The variance of photon number is found to be
\begin{align}
\langle(\Delta\hat{n}^{\rm{w}}_{b,L})^2\rangle 
= |p|^4\langle(\Delta\hat{n}^{\rm{w}}_{b})^2\rangle + |p|^2|q|^2\langle\hat{n}^{\rm{w}}_{b}\rangle.
\end{align}
The average Fano factor can be then obtained
\begin{align}
\overline{F_{L}}=&\dfrac{\overline{\langle(\Delta\hat{n}^{\rm{w}}_{b,L})^2\rangle}}{\overline{\langle\hat{n}^{\rm{w}}_{b,L}\rangle} }
= |p|^2 \overline{F} +|q|^2.
\end{align}
The average SNR arrives at
\begin{align}
\frac{\overline{R_{L}}}{\overline{\langle \hat{n}_{b,L}^{\rm{w}}\rangle}}=1/\overline{F_{L}}.
\end{align}


\section*{Acknowledgements}


We would like to thank Prof. S. Sun for valuable discussions. This work was supported by Science Challenge Program (TZ2018003-3) and National Natural Science Foundation of China (NSFC) (61875178, 11605166).

\section*{Author contributions statement}


All authors contributed to the numerical investigations, the discussions of the results and the preparation of the manuscript.


\section*{Additional information}

\textbf{Competing interests:} The authors declare no competing interests.






\end{document}